\documentclass[
 reprint,
superscriptaddress,
nofootinbib,
nobibnotes,
 amsmath,amssymb,
 aps,
]{revtex4-2}
\usepackage{graphicx}
\usepackage{dcolumn}
\usepackage{bm}
\usepackage{xcolor}
\usepackage{amsmath}	
\setcitestyle{citesep={,}}

\usepackage{hyperref}
\usepackage[mathlines]{lineno}

\newcommand{\colore}{\texttt{CoLoRe}}
\newcommand{\be}{\begin{equation}}
\newcommand{\ee}{\end{equation}}
\newcommand\bea{\begin{eqnarray}}
\newcommand\eea{\end{eqnarray}}

\newcommand\nside{\texttt{NSIDE}}
\newcommand\Ntpl{N_{\mathrm{tpl}}}

\newcommand\Cl{C_\ell}

\newcommand{\addfrac}{\eta_{\rm add}}
\newcommand{\threetimestwo}{3$\times$2}

\newcommand{\DNF}{\ifmmode \mathtt{DNF} \else \texttt{DNF} \fi}

\newcommand{\xvec}{\Vec{x}}

\newcommand{\fmult}{f_{\rm mult}}
\newcommand{\fadd}{f_{\rm add}}

\newcommand{\fmultp}{f'_{\rm mult}}

\newcommand{\deltasub}[1]{\delta_{\rm{#1}}}
\newcommand{\Nsub}[1]{N_{\rm{#1}}}

\newcommand{\maglimpp}{\textsc{MagLim}\texttt{++}}
\newcommand{\decals}{\texttt{DECaLS}}
\newcommand{\unWISE}{\texttt{unWISE}}
\newcommand{\WISE}{\texttt{WISE}}
\newcommand{\wise}{\texttt{WISE}}
\newcommand{\maglim}{\texttt{MagLim}}
\newcommand{\redmagic}{\texttt{RedMaGiC}}
\newcommand{\dnf}{\texttt{DNF}}

\newcommand{\zmean}{\texttt{ZMEAN}}
\newcommand{\extmash}{\texttt{EXTMASH}}

\newcommand{\healpix}{\textsc{HealPix}}

\begin{document}

\preprint{APS/123-QED}

\title[Bin-optimized star-galaxy separation]{Taming additive systematics via redshift-bin-optimized star-galaxy separation}

\author{Noah Weaverdyck}
\email{NWeaverd@proton.me}
 \affiliation{Lawrence Berkeley National Laboratory, 1 Cyclotron Road, Berkeley, CA 93720, USA}
 \affiliation{Berkeley Center for Cosmological Physics, Department of Physics, UC Berkeley, CA 94720, USA}

\author{David Schlegel}
 \affiliation{Lawrence Berkeley National Laboratory, 1 Cyclotron Road, Berkeley, CA 93720, USA}

 \author{Anand Raichoor}
 \affiliation{Lawrence Berkeley National Laboratory, 1 Cyclotron Road, Berkeley, CA 93720, USA}

\author{Ignacio Sevilla-Noarbe}
\affiliation{Centro de Investigaciones Energ\'eticas, Medioambientales y Tecnol\'ogicas (CIEMAT), Madrid, Spain}


\date{\today}

\begin{abstract}
Contamination from stars in the galaxy samples of large-scale structure surveys can bias cosmological constraints if not tightly controlled. This is especially true for lens samples used for galaxy clustering and galaxy-galaxy lensing probes, where contamination is a primary source of additive systematics. We propose an improved approach to star-galaxy separation and an optimal weighting scheme to jointly mitigate additive and multiplicative contamination of the density field at the map level.
Our star-galaxy separation approach exploits the fact that photometric galaxy samples used for cosmological inference populate different regions of color-space than the full photometric dataset on which star-galaxy cuts are typically applied, and therefore optimizes star-galaxy separation for the galaxy samples in each redshift bin. 
This serves as a complementary approach to morphological star-galaxy separators, which can have complicated dependencies on PSF and blending systematics. We demonstrate the method using the Dark Energy Survey Y3 \maglim{} lens sample, for which we obtain forced NIR unWISE photometry via cross-matching with \decals{} DR9 to define redshift-bin-optimized color cuts. We identify and remove residual stellar contamination in the DES Y3 lens sample, which varies strongly across redshift bins ($1.3-5.5\%$) and across the footprint.
\end{abstract}

\maketitle


\section{Introduction}
Large-scale structure (LSS) surveys have become one of the most powerful probes of cosmology, enabling precise measurements of dark energy, dark matter, and the growth of structure in the Universe. Modern photometric galaxy surveys such as the Dark Energy Survey \citep[DES;][]{DES:2005}, the Kilo-Degree Survey \citep[KiDS;][]{2013ExA....35...25D}, and the Hyper Suprime-Cam Subaru Strategic Program \citep[HSC;][]{2022PASJ...74..247A} have mapped the locations and shapes of hundreds of millions of galaxies across thousands of square degrees, with upcoming surveys like the Vera C. Rubin Observatory's Legacy Survey of Space and Time \citep[LSST;][]{Ivezic:2019} and Euclid \citep{2025A&A...697A...1E} poised to increase these numbers by orders of magnitude.

A wealth of cosmological information is embedded in these maps of LSS, but to extract it typically requires compressing them into summary statistics that can be compared against theoretical predictions. 
Two-point statistics have been the primary observables used in the flagship analyses of these large surveys, with the so-called \threetimestwo{}pt analysis combining information from both the galaxy density and shear fields. Two-point statistics capture most of the cosmological information content up to the quasi-linear regime where the fields can be treated as Gaussian, but are insensitive to the non-Gaussian information present.
A large number of additional summary statistics have been proposed for this purpose, including higher-order statistics \citep{Takada_2003, Schmittfull_2015}, $k$-NN statistics \cite{Banerjee_2020}, skew spectra \cite{Dizgah_2020}, marked spectra \cite{White_2016, Yin_2025}, wavelet and scattering transforms \citep{Cheng__2020, Valogiannis_2022}, and more. However, all of these are ultimately derived from the underlying galaxy density fields, such that clean catalogs and maps are crucial to avoid hard-to-model systematics from biasing cosmological inference, and far preferable to \textit{post-facto} hardening of each statistic to field-level contamination.

A persistent challenge in photometric surveys is the presence of spurious density fluctuations because of contamination from other sources and variation in the so-called selection function of galaxies, which characterizes how the probability of detecting a given galaxy varies across the footprint. These spurious fluctuations are variously termed calibration errors, or angular/imaging/LSS systematics. Reducing, characterizing and mitigating these spurious fluctuations is a key component of cosmological analyses using galaxy clustering statistics \cite{Huterer_2013, Weaverdyck:2020mff, Kitanidis_2020, DES:2021bat, Rezaie_2024, Yan_2025}.

As argued in \cite{weaverdyck2026darkenergysurveyyear} and \cite{rodriguezmonroy2025darkenergysurveyyear}, while sample definition, footprint definition, and residual angular systematics corrections often occur in a disconnected and somewhat independent fashion, it is preferable for these processes to inform one another, as the process of correcting for systematics can simultaneously help to identify limitations of such \textit{post-facto} corrections, including regions of the sky or color-space which would be better removed altogether.
This work is motivated by one such case: the challenge of separating and correcting for the multiple competing effects of stars on the observed galaxy density and summary statistics used for cosmological inference.

The standard approach to mitigating angular systematics in galaxy surveys is to use maps that trace local properties as spatial templates for contamination, and to remove clustering corresponding to those spatial patterns either from the two-point statistics or from the density field directly. Stellar density, dust extinction, depth and PSF size are all common templates used for these purposes. Different methods make different, often tacit assumptions about the form of systematic contamination and the true overdensity field \cite{Weaverdyck:2020mff}.

While most systematics are multiplicative (i.e. they modulate the observed density in a multiplicative fashion), there are also  additive systematics which are independent of the true density field, and treating both of these simultaneously at the map level is nontrivial. 
One key difference is that multiplicative systematics couple small and large scales together \cite{Shafer_2015}, in effect inducing a spatially dependent noise term in the observed galaxy density \cite{Xavier_2019, Weaverdyck:2020mff}. Several recent works have proposed methods for differentiating multiplicative from additive systematics using this dependence. \citet{Xavier_2019} focus on stars as the main additive contaminant and propose a multi-step process of using the variance of the observed over-density to first estimate and correct for stellar obscuration of background galaxies (a multiplicative effect) and then subsequently use the spatial dependence of the mean to estimate and correct for stellar contamination. 
More recently, \citet{Berlfein:2024uwi} proposed to jointly fit and mitigate additive and multiplicative systematics for all templates at the two-point level, but at the cost of additional suppression of true LSS signal 
and without a map-level estimator for the case where both are present.
\citet{Hern_ndez_Monteagudo_2025} also propose a multi-stage approach applied to a full list of templates, where they alternate nulling any significant dependencies they find of the mean and variance of the observed number density against their template library. 

In the first part of this work, we derive a map-level overdensity estimator for both additive and multiplicative systematics, which fully accounts for the impact of additive systematics on the integral constraint, which has previously been neglected. 

However, we also argue that rather than attempting to correct for stellar contamination through weighting schemes, direct removal of contaminating sources provides a more robust solution. To this end, the second and main part of this work describes an optimized star-galaxy separation method that leverages the distinct regions of color-space occupied by different redshift bins of cosmological galaxy samples.
By tailoring color cuts to each redshift bin of the galaxy sample, we produce a purer sample as compared to traditional morphology-based separators, which suffer from complex dependencies on PSF variations and blending, or color-based selections that do not account for how star-galaxy confusion varies with photometric redshift. 

The rest of this paper is organized as follows. In Section 2, we present the theoretical framework for additive and multiplicative systematics and derive a generalized galaxy weight to treat both of these at the map-level. We demonstrate its performance on a simulated galaxy sample contaminated with stars. In Section 3 we describe our improved method for star-galaxy separation conditioned by redshift bin and apply it to the DES Y3 \maglim{} sample. Section 4 describes our results identifying residual stellar contamination in the \maglim{} sample. We discuss the implications of our findings and conclude in Section 5.

\section{Additive and multiplicative systematics}\label{sec:addmult}
Stars can affect cosmological inference with photometric surveys in multiple and competing ways. For instance, stars can be misclassified as galaxies in the sample (``contamination"), or the presence of bright stars can make it harder to detect nearby galaxies (``obscuration" or ``occultation") because bright wings make object detection more difficult or because they result in overestimation (and subsequent subtraction) of the local sky background \cite{Ross_2011, Kalus_2018}. While contamination adds to the observed density field in a manner that is independent of the actual galaxy density field to first order, obscuration instead results in a \textit{relative} suppression of the observed galaxy density field. 

While our primary motivation is to deal with these effects from stars, they apply generically to any additive and multiplicative systematics and so we use general notation; specialization to the case of stars can be achieved via the substitution $\fmult \rightarrow f_{\rm obscuration}$ and $\fadd \rightarrow f_{\rm contamination}$. 
One can show (App.~\ref{sec:app1}) that the combination of unmitigated additive and multiplicative systematics results in the observed overdensity field ($\deltasub{obs}$) being a biased estimate of the truth ($\deltasub{true}$)\footnote{Note that \citet{Berlfein:2024uwi} assume this same form but miss the prefactor from the integral constraint in front of $\deltasub{true}$ in their Eq.~2, which can be rather important as we'll show.}: 

\begin{align}
\deltasub{obs} &\equiv \frac{N_{\rm obs}}{\overline{N}_{\rm obs}} - 1 \\
                &= \left(1 - \frac{\overline{N}_{\rm add}}{\overline{N}_{\rm obs}}\right) \deltasub{true} + \fmult + \fadd + \fmult \deltasub{true}, \label{eq:deltaobs}
\end{align}
where an overbar indicates the spatial average over the sample (e.g. redshift bin), and other terms have an implicit dependence on both angle and redshift. We have used $N_{\rm obs} \equiv \Nsub{true} + \Nsub{add} + \Nsub{mult}$ to indicate the observed density of objects, which includes density contributions from true galaxies $(\Nsub{true})$, interloping objects that are independent of the true galaxy field $(\Nsub{add})$, and interloping objects (or when negative, missing true galaxies) with a scalar dependence on the true density field $(\Nsub{mult})$. 
As shown in App.~\ref{sec:app1}, these can be written as density contrasts with respect to the observed density, $\fadd$ and $\fmult$ respectively, which are usually estimated from the data using systematic templates intended to characterize their spatial dependence.

Assuming $\fmult$ and $\fadd$ do not themselves trace LSS, then terms with $\langle \deltasub{true} f_{\rm x}\rangle$, $\langle f_{\rm x} (f_{\rm x} \deltasub{true})\rangle$ or $\langle \deltasub{true} (f_{\rm x} \deltasub{true})\rangle$ vanish on average, giving the resultant two point function:
\begin{equation}
\begin{split}
\langle\deltasub{obs}\deltasub{obs}\rangle &= \left(1 - \frac{\overline{N}_{\rm add}}{\overline{N}_{\rm obs}}\right)^2 \langle \deltasub{true}\deltasub{true}\rangle \\
&\quad + \langle\fmult\fmult\rangle + \langle\fadd\fadd\rangle + 2\langle\fmult\fadd\rangle \\
&\quad + \langle(\fmult \deltasub{true})(\fmult \deltasub{true})\rangle
\end{split}
\label{eq:twopt_obs}
\end{equation}

From Eqs.~\ref{eq:deltaobs} and \ref{eq:twopt_obs} we see a few effects:
\begin{enumerate}
    \item A constant multiplicative suppression of the galaxy two point function due to including spurious additive contaminants when computing the normalization, i.e. the wrong integral constraint. This is a known effect, see e.g. \citep{Crocce_2015, Nicola_2020, Krolewski_2020} who show the impact of residual stellar contamination in the sample, in the limit that such contamination is uniform across the footprint (i.e. unclustered, with the fluctuation ${f}_{\rm add} = 0$).
    As noted in the latter, the impact is degenerate with linear galaxy bias for both auto and cross-power statistics (if one assumes no other systematics).
    We note, however, that the effective bias will change when using a footprint with a different stellar contamination rate, such as when cross-correlating a subarea of the footprint with another probe, or when performing area-split tests. Both of these cases may show unexpected inconsistencies with the main analysis due to additive contamination levels that vary with area, manifesting as a different preferred effective galaxy bias. The DES Y3 clustering analysis did not explicitly infer a residual mean stellar contamination fraction, thus absorbing it into the linear galaxy bias parameters.  
    \item \textit{Both} additive and multiplicative systematics contribute an additive term to the overdensity: $\fadd$ and $\fmult$. If both trace a common spatial template, then these are degenerate when using standard regression-based approaches with the observed overdensity (note that with stars, obscuration and contamination impact at this order will partially cancel). 
    \item Multiplicative systematics contribute a coupling term that is multiplicative with the true overdensity. As shown in \citep{Weaverdyck:2020mff}, this effectively imprints a spatially-dependent \textit{variance} in the observed overdensity field in the pattern of the given systematic, which propagates to any higher N-point functions of the field, such as the last coupling term in Eq.~\ref{eq:twopt_obs}, and can have significant impacts if not adequately addressed \citep{Shafer_2015, Weaverdyck:2020mff, Berlfein:2024uwi}. 
\end{enumerate}

Treating systematics as only additive (as with pseudo-$\Cl$ Mode Deprojection \citep{Elsner_2016, Alonso_2019, Nicola_2020, Cornish_2026}) or as only multiplicative via galaxy weights (as in \citep{DES:2021bat, desicollaboration2024desi2024iisample}) only effectively treats the associated additive or multiplicative terms, such that summary statistics will be biased in the case where both types of systematics are present \citep{Shafer_2015, Weaverdyck:2020mff, Berlfein:2024uwi}. 

In Ref.~\cite{Weaverdyck:2020mff} we showed how standard regression tools (which implicitly assume additive contamination) can be applied and then corrected to address multiplicative contamination, but this requires the additive and multiplicative systematics to have different and known templates, and did not address the map-level bias on the overdensity field coming from the integral constraint described in bullet (1).

We therefore derive a general estimator of the true overdensity field in the presence of both additive and multiplicative systematics, as well as corresponding field-level weights to produce accurate summary statistics from said fields ($N$-pt functions, scattering transforms, skew spectra, $k$-NN statistics, etc.).

Using the notation of Eq.~\ref{eq:deltaobs} an unbiased estimate of the true overdensity field can be written as
\be
\hat{\delta}_{\rm true} \approx \frac{\delta_{\rm obs} - {f}_{\rm{add}} -  {f}_{{\rm mult}}}{1 + {f}_{\rm mult} - \addfrac},\label{eq:deltaest}
\ee
where $\addfrac = \overline{N}_{\rm add}/\overline{N}_{\rm obs}$ is the total fractional additive contamination and $\hat{X}$ indicates an estimate of $X$. From here on, we'll use $\hat{\delta}\equiv \hat{\delta}_{\rm true}$ for brevity.

\begin{figure*}
    \centering
    \includegraphics[width=1\linewidth]{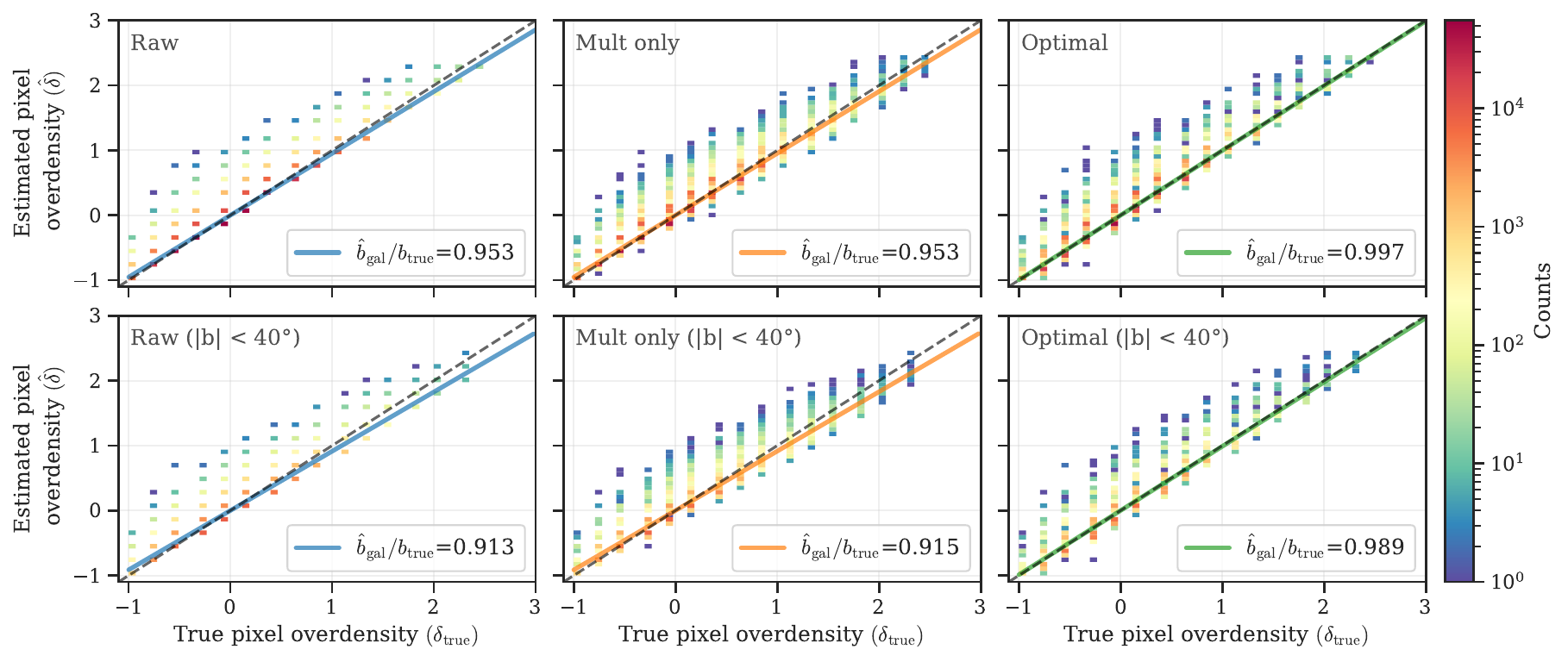}
    \caption{Comparison of the estimated and true overdensity in \nside{} 512 Healpixels for a simulated galaxy catalog with 5\% stellar contamination using the  raw observed field with no weights (left), multiplicative-only weights (Eq.~\ref{eq:fullmultweight}, center) and the optimal weights from Eq.~\ref{eq:optweight} (right). The top row compares the full map, whereas the bottom row uses the same galaxy weights computed from the full-map, but computes the overdensity for the $\sim20$\% of the footprint at low galactic latitude ($|b|<40^\circ$). Even though the weights remove the spatial correlation with stellar density (and its effects on the 2-point function, c.f. Fig.~\ref{fig:wtheta_sim}), suppression of the overdensity remains due to the wrong normalization for the integral constraint. The colored line indicates the fitted amplitude of estimated to true overdensity, and the slope reported in the legend gives the ratio of the inferred vs. true galaxy bias: $\hat{b}_{\rm gal}/b_{\rm true}$. Failure to treat the integral constraint term (which requires a robust estimate of the average contamination rate) results in a biased estimate of the galaxy bias. Using a subset of the data as shown in the bottom row (e.g. for cross-correlation with another survey) will result in a \textit{different} and also incorrect, effective bias, because the actual stellar contamination rate varies strongly across the footprint. With the optimal weights in the right column, we recover the correct galaxy bias for both the full and subsample.}
    \label{fig:delta_compare}
\end{figure*}

Optimal galaxy weights\footnote{Of course where there are no galaxies, no galaxy weight can correctly account for the selection function, which illustrates how modulating the \textit{randoms} as $1/w_g$ is really the more correct thing to do and also properly accounts for the impact on the covariance (see e.g. \cite{Morrison_2015}). This also addresses the issue of $w_g$ diverging as $\deltasub{obs}\rightarrow-1$.} that account for both the multiplicative and additive effects can be computed generally as

\begin{align}
w_g^{\rm opt}  &= \frac{1 + \hat{\delta}}{1 + \deltasub{obs}},\label{eq:weight}\\
		  &= \frac{1 +  ({\delta_{\rm obs} - \hat{f}_{\rm{add}} -  \hat{f}_{{\rm mult}}})/(1 + \hat{f}_{\rm mult} - \hat{\eta}_{\rm add})}{1 + \deltasub{obs}} \label{eq:optweight},
\end{align}
where estimates of the (zero-mean) contamination fields, $\hat{f}_X$, can come from a variety of methods that can largely be recast as 
some form of regression of $\delta_{\rm obs}$ against a set of template maps \cite{Weaverdyck:2020mff}. 

Eq.~\ref{eq:optweight} thus provides, for the first time to our knowledge, an explicit prescription for galaxy weights that correct for \textit{both} additive and multiplicative systematics in a galaxy sample at the map level.

To fully treat only additive systematics at the map level (including the integral constraint), Eq.~\ref{eq:optweight} reduces to 
\be
w_g^{\rm add} = \frac{1 - (\addfrac + \hat{f}_{\rm add})/(1+{\delta_{\rm obs}})}{1-\addfrac},\label{eq:fulladdweight}
\ee
and for the multiplicative-only case to the familiar 
\be
w_g^{\rm mult} = 1/({1+\hat{f}_{\rm{mult}})}.\label{eq:fullmultweight}
\ee

\subsection{Testing impact of different weighting schemes on galaxy clustering}
We test the impact of stellar contamination and mitigation via different weighting schemes by simulating and cleaning a galaxy catalog with a known level of stellar contamination. For our simulated catalog, we adopt the best-fit DES Y3 cosmology \cite{Abbott_2022}, and linear galaxy bias $b$, number density $\bar{n}$, and redshift distribution $n(z)$ \cite{Cawthon_2022} for the third redshift bin from the fiducial DES Y3 \maglim{} sample \cite{Porredon_2022}, which we use as input to \colore{} \cite{Ram_rez_P_rez_2022}. More detail on the \maglim{} sample is given in Sec.~\ref{sec:catalog_rejection}, but is not relevant for our simulated tests here, for which we only need a catalog that is reasonably representative of those used for galaxy clustering constraints.

For simplicity, our simulated footprint has only a coarse mask consisting of all \texttt{NSIDE} 512 pixels with any contribution to the actual DES Y3 footprint (that is, we do not bother simulating holes finer than this resolution). We then generate 25 million randoms in this footprint for each redshift bin, i.e. $\sim$20$\times$ the number of galaxies.

We simulate a sample of $\addfrac \times N_{\rm gal}$ contaminating stars by Poisson sampling from an \texttt{NSIDE} 512 stellar density template derived from Gaia observations and described in \cite{rodriguezmonroy2025darkenergysurveyyear}. We use $\addfrac=0.05$, which is a reasonable prior for a single bin in current surveys. This produces an average of 0.22 stars per 512 \healpix \cite{Gorski_2005} pixel (\textit{Healpixel}), but with an integer number that is observed and uniformly scattered to span the full area of the pixel. 
We then estimate the contamination field $f$ by regressing the observed overdensity field against our stellar density template\footnote{This assumes that $\deltasub{obs} = \deltasub{true} + f + \epsilon = \hat{\delta} + \hat{f}$, which as noted in \cite{Weaverdyck:2020mff} gives an unbiased estimate for $f$ regardless of whether the true contamination is additive or multiplicative. However, any portion of $f$ or $\epsilon$ that is not fit by the contamination model will tacitly be absorbed into $\hat\delta$, including deviations between the mean and observed contamination due to e.g. Poisson noise in $\epsilon$.}
$t_{\rm star}$:
$\hat{f}=\hat{\alpha}t_{\rm star}$ where $\hat{\alpha} = {\rm argmin}_\alpha ||\deltasub{obs} - \alpha t_{\rm star}||^2$. From this we generate standard multiplicative weights (Eq.~\ref{eq:fullmultweight}) as well as the optimal from Eq.~\ref{eq:optweight} (which in this case is equivalent to Eq.~\ref{eq:fulladdweight}).

\subsubsection*{Map-level comparison}
We first check the fidelity of map-level estimates with the different weights. 
Fig.~\ref{fig:delta_compare} compares the estimated and true galaxy overdensity (unsmoothed, in Healpixels of \nside{} 512) for a simulated catalog with 5\% stellar contamination. Different columns have different weights applied (unweighted, multiplicative, and optimal) and the top vs. bottom rows are for the full vs. a subset of the footprint. 
Discreteness in $\deltasub{true}$ (as well as in $\hat\delta$ for the raw, unweighted case), arises due to the integer object counts per pixel. At these scales, the pixel-level error is dominated by the Poisson error in the stellar contamination,\footnote{Even with this relatively large contamination rate of $5\%$, we have an average number of $\bar{n}_{\rm star}\approx 16.3$ deg$^{-2}$ such that even though our stellar template traces the mean number of stars across the footprint, scales below $\sim0.25^\circ$ (corresponding to $\nside{} \gtrsim 256$) are dominated by Poisson noise.} $\sigma_{\delta,\ \mathrm{Pois}} = \sqrt{\eta_{\rm add}/\langle{N}_{\rm obs}\rangle_{\rm pix}}$ such that regardless of how or whether the dependence on stellar density is removed via weights, the overdensity error is RMSE($\delta$) $= \sigma_{\deltasub{true} - \hat{\delta}} \approx 0.11$. A similar comparison at \nside=128 gives the same bias results, but RMSE = $\{0.043, 0.031, 0.031\}$, for the raw, multiplicative and optimal weights cases on the full footprint, respectively, showing how the weights improve the typical map-level error as Poisson noise is reduced ($\sigma_{\delta_{128},\ \mathrm{Pois}}\approx \sqrt{0.05/72.5}=0.026$).  

While the map-level RMSE shows little difference between the multiplicative and optimal weights, we can identify the difference in the inferred galaxy bias noted in the previous section when discussing Eq.~\ref{eq:twopt_obs}. Failing to account for the mean contamination rate and its impact on the integral constraint results in a suppression of the inferred galaxy bias, which is demonstrated by the slope by the best-fit line comparing the amplitudes of $\hat\delta$ to $\deltasub{true}$ in each subplot ($\hat{b}_{\rm gal}/b_{\rm true} = \langle\hat\delta \deltasub{true}\rangle/\langle\deltasub{true}\deltasub{true}\rangle$). 

Crucially, this bias is \textit{not} uniform across the footprint: the bottom row of Fig.\ref{fig:delta_compare} shows the same comparison, but for a subset of the footprint. The galaxy weights are the same as for the full catalog, but the overdensity is recomputed only considering the footprint at low galactic latitude ($|b|<40^\circ$), as is common when e.g. cross-correlating a galaxy sample with another probe, such as CMB lensing. In this region, the stellar contamination rate is higher, and so the effective linear galaxy bias for the sample differs from that inferred for the catalog on the full footprint. The optimal weights, which account for the integral constraint, recover the correct galaxy bias on both the full footprint and sub-footprint, with $\hat{b}_{\rm gal}/b_{\rm true} \approx 1$.

\subsubsection*{2-pt comparison}
Next, we compare the fidelity of the recovered 2-pt function with the different weights, as this is the quantity used most often for cosmological inference.
We use the \texttt{treecorr} \cite{Jarvis_2004} \texttt{NNcorrelation} function to compute the two-point angular auto-correlation $w(\theta)$ of the observed sample via the standard Landy-Szalay estimator \cite{landy_szalay} using different sets of galaxy weights and compare them to that computed on the catalog with no stellar contamination.\footnote{For \texttt{treecorr} settings, we use \texttt{binslop} = 0.02 to compute $w(\theta)$ in 30 log-spaced bins for $\theta \in [2.5,\ 2500]$ arcmin. For the covariance, we use 400 patches generated via \texttt{treecorr} from the randoms and \texttt{cross\_patch\_weight = 'match'} to use the jackknife correction from \cite{Mohammad_2022}. We confirmed results don't depend sensitively on these choices.}

\begin{figure*}
    \centering
    \includegraphics[width=.8\textwidth]{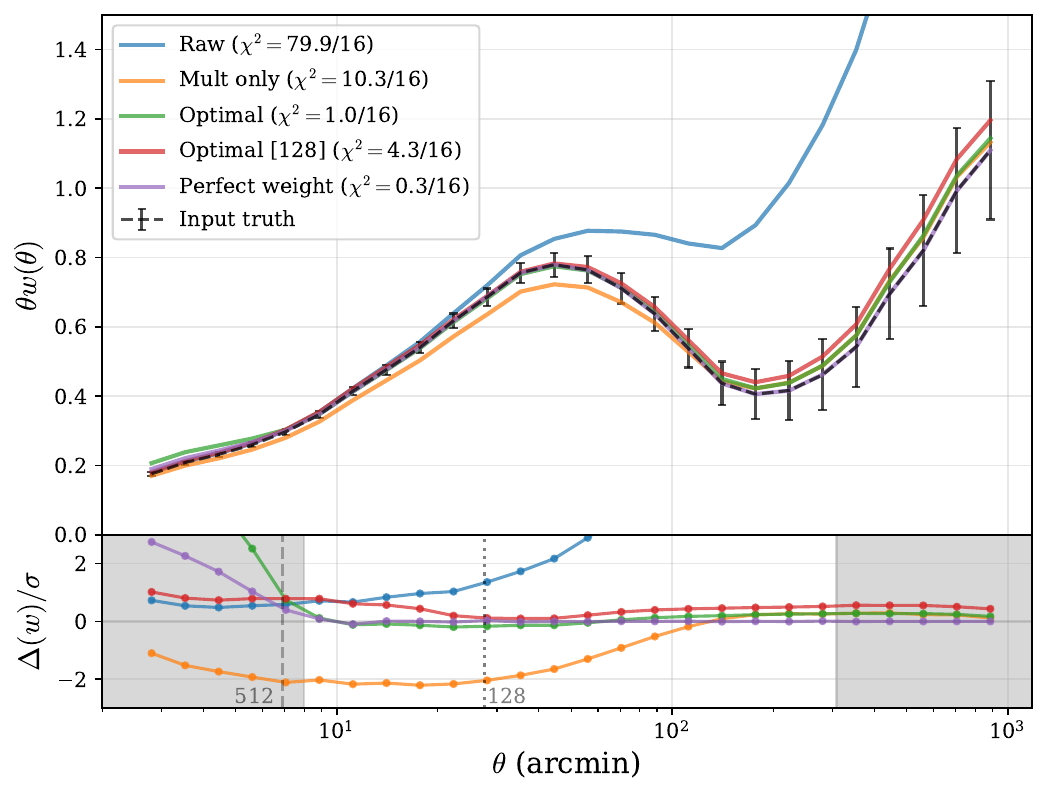}
    \caption{Angular correlation function $w(\theta)$ for redshift bin 3 of a simulated DES Maglim sample using \colore\cite{Ram_rez_P_rez_2022}. The input  $w(\theta)$ is given in black, with jackknife error bars. We then contaminate the sample with 5\% stellar contamination, and show $w(\theta)$ for the raw sample (blue), and after the standard multiplicative-only weights from Eq.~\ref{eq:fullmultweight} are applied (orange) or the more optimal weights accounting for additive systematics via Eq.~\ref{eq:fulladdweight} are applied (green), assuming a known total stellar contamination rate. 
    The bottom subplot shows the difference with respect to the input truth, normalized to the uncertainty, with the $\chi^2$ difference given in the legend for scales not removed by scale cuts for \threetimestwo{}pt analyses \cite{DES:2021bat} (indicated in gray). 
    Using multiplicative-only weights with this additive contamination effectively removes contamination at large scales, but is biased at small to medium scales. Most of this bias is due to the integral constraint term described in Sec.~\ref{sec:addmult}, and is almost fully removed using the optimal weights described in Eq.~\ref{eq:fulladdweight}. Estimating the weights at a lower resolution of \texttt{NSIDE}=128 (red), results in residual bias below the pixel scale (indicated via dotted [dashed] line for \texttt{NSIDE}=128 [512]), but still an improvement over the multiplicative-only case. 
    In purple we show the ``perfect" weights case, where we know \textit{a priori} at a pixel level how many objects are stars, and thus apply weights as $w_g^i=\frac{N^i_{\rm true}/\bar{N}_{\rm true}}{N^i_{\rm obs}/\bar{N}_{\rm obs}}$, 
    bypassing the stellar density template. 
    }
    \label{fig:wtheta_sim}
\end{figure*}

Fig.~\ref{fig:wtheta_sim} shows $w(\theta)$ for the third redshift bin of our simulated sample. In black with jackknife errors is the true (uncontaminated) $w(\theta)$. In blue is the same sample but with 5\% stellar contamination added, with the corresponding normalized residual with respect to the truth shown in the bottom subplot. This raw contaminated sample shows significant deviation from the true correlation function, with $\chi^2/\nu_{\rm dof}=80/16$, where 
\be
\chi^2 = (w_{\rm X} - w_{\rm true})^T \textbf{C}^{-1} (w_{\rm X} - w_{\rm true}),
\ee
and $\textbf{C}$ is the jackknife covariance of $w_{\rm true}$ for the 16 data points that fall within $8 < \theta < 300$ arcmin.

If we treat it as multiplicative and apply the standard multiplicative weights from Eq.~\ref{eq:fullmultweight}, we effectively remove the large-scale contamination but obtain biased $w(\theta)$ at small to medium scales (orange). This bias primarily arises from the integral constraint term discussed in Section \ref{sec:addmult}. 
If we assume we can accurately determine the total stellar contamination rate $\addfrac$, then we can  accurately treat the contamination as additive via the weights in Eq.~\ref{eq:fulladdweight} (green) and all but eliminate the systematic bias, recovering well the input correlation function down to the approximate 512 pixel scale (dashed line at $\sim7$ arcmin), though estimates diverge below this.

We also test the impact of computing the weights at \texttt{NSIDE}=128 rather than \texttt{NSIDE}=512 in red. This is because there's a tradeoff -- lower resolution gives pixelized galaxy densities that are better described as Gaussian and with less extreme discreteness effects from the integer galaxy counts, and the corresponding additive weights estimates are more stable. However, these weights will fail to correct for systematic contamination below the resolution scale. We find that $w(\theta)$ with the 128 additive weights does show significant bias below the 128 pixel scale (dotted at $\sim28'$), but is still an improvement over the multiplicative-only approach at 512.

For comparison, we also include the idealized case where the exact number of contaminating stars in each 512 Healpixel is known \textit{a priori}, allowing direct calculation of weights as $w_g^i = (N^i_{\rm true}/\bar{N}_{\rm true})/(N^i_{\rm obs}/\bar{N}_{\rm obs})$ without requiring a stellar density template. This perfect-knowledge scenario (purple) achieves $\chi^2 = 0.3$ and demonstrates the theoretical limit for using weights at this resolution.

\subsection{Challenges}\label{sec:challenges}

While the optimal weighting scheme described above clearly improves upon the multiplicative-only case when there are additive contaminants like stars, there are a few challenges:
\begin{enumerate}
    \item An independent estimate of the fractional contamination $\addfrac$ is required. $\hat\eta_{\rm add}$ has been estimated by e.g. using deep fields that contain more data \cite{Nicola_2020}, simulated source injection \cite{Yamamoto_2025, Anbajagane_2025}, or extrapolating observed trends of $N_{\rm gal}$ against stellar density to where the latter is zero \cite{Crocce_2018, Rosell_2021}. Each of these has their own challenges; stellar contamination or density estimates from deep field observations or injections must be extrapolated to the full footprint, accounting for the fact that stellar density has strong spatial dependence (something not always done in practice). Extrapolating the observed galaxy density to where the stellar density is zero will underestimate the true contamination, as it assumes no impact of stellar obscuration.
    \item $\hat{f}_{\rm{add}}$ and $\hat{f}_{\rm{mult}}$ are both typically obtained by regressing the observed overdensity against spatial templates, such that if a multiplicative and additive systematic both have the \textit{same} spatial template (as is the case with stellar contamination and obscuration), then more complex methods are required to break the degeneracy as in e.g. \citep{Xavier_2019, Berlfein:2024uwi, Hern_ndez_Monteagudo_2025};
    \item Systematic templates are only approximate tracers of the true systematic, such that corrections will always be approximate. While a template comprised for e.g. Gaia star density will remove the mean dependence on stellar density, it can't correct for the Poisson residuals, or fluctuations below the resolution at which its computed. Furthermore, it assumes that contamination and obscuration rates are independent of other features such as star color and magnitude, which may or may not be true in practice.
\end{enumerate}

These challenges motivate direct removal of additive contaminants like stars at the catalog level where possible, as this sidesteps the need to model and jointly fit additive and multiplicative systematics sharing the same spatial templates, and avoids sacrificing hard-won cosmological signal.

\section{Improving stellar rejection at the catalog level}\label{sec:catalog_rejection}
For the rest of this work, we describe how we can remove the dominant additive systematic in galaxy samples by leveraging near infrared (NIR) data to optimize stellar rejection \textit{accounting for the color-distributions in each redshift bin}, as a means to complement and improve upon the sample-wide morphological star-galaxy classifiers like those used in the DES Y3 \maglim{} lens galaxy catalog \citep{2021PhRvD.103d3503P}.

\subsection{Original DES Y3 \maglim{} sample}
Cosmological constraints from DES Y3 used two different lens galaxy samples with quite different selection methodologies. Due to unresolved systematics in the \redmagic{} catalog \citep{2016MNRAS.461.1431R}, which used a photometric template to select LRGs, the much larger \maglim{} catalog \cite{2021PhRvD.103d3503P} was ultimately selected as the fiducial \cite{DES:2021bat, Pandey_2022}. 
\maglim{} is selected from the Y3 Gold catalog using objects that pass certain quality cuts and satisfy 
a photo-z dependent \textit{i}-band magnitude limit:
\begin{equation}
    17.5 < i < 18 + 4 \times \zmean{},~~~ \zmean \in [0.2, 1.05],\label{eq:maglim_select}
\end{equation}
where \zmean{} is the photometric redshift predicted from a linear fit in \textit{griz} color space using 80 nearest neighbors and a distance metric that prioritizes color differences over overall flux differences (c.f. the Directional Neighbourhood Fitting (\DNF) algorithm, \citep{2016MNRAS.459.3078D}). 
Objects are then binned into six redshift bins according to their \zmean{} with edges $[0.2, 0.4, 0.55, 0.70, 0.85, 0.95, 1.05]$. 
Finally, only objects that are galaxies with high confidence are retained, via the selection \extmash{}\footnote{We use the shorthand \extmash{} to indicate \texttt{EXTENDED\_CLASS\_MASH\_SOF}.} = 3, which selects galaxies that are extended with high confidence \cite{Sevilla_Noarbe_2021}.

\begin{figure*}
    \centering
    \includegraphics[width=1\linewidth]{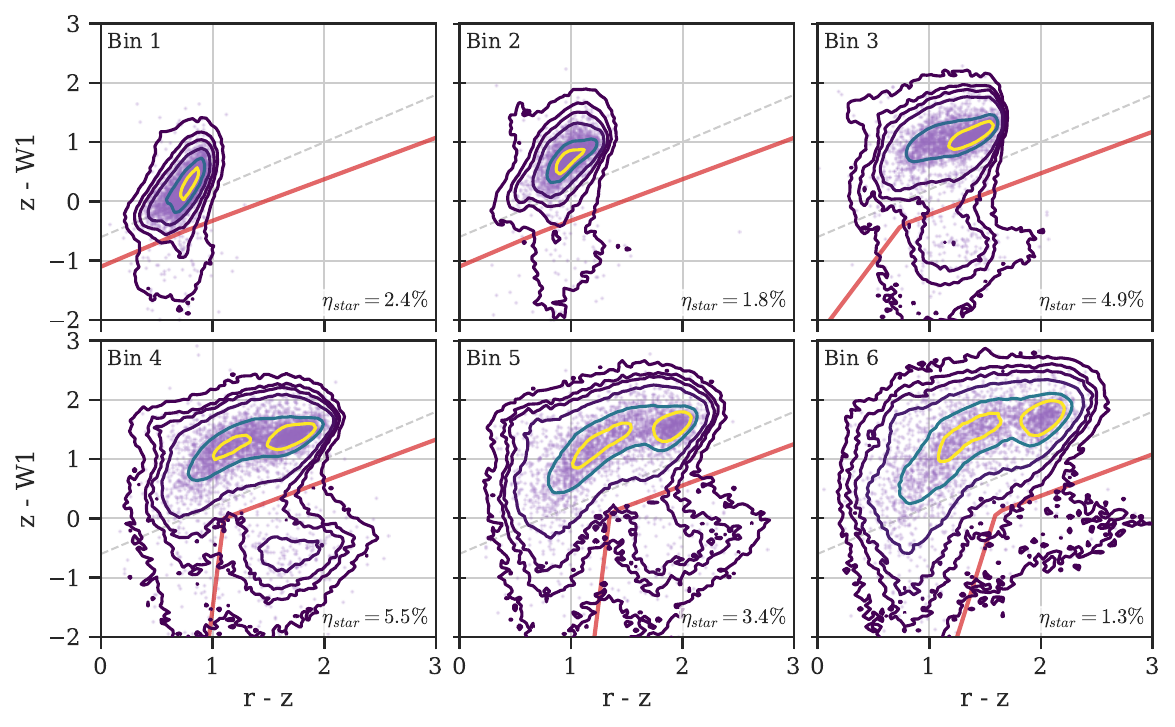}
    \caption{The distribution of objects in the full DES Y3 Maglim sample in each redshift bin. Points are downsampled by 500 and contour lines give the $\{32, 68, 90, 95, 97, 99\}^{\rm th}$ \%-iles of the full distribution. The broken red line indicates the redshift-bin-optimized star-galaxy separator, which isolates the island of residual contaminating stars in each bin, with the total fraction of removed stars indicated by $\eta_{\rm star}$. The dashed gray line shows the redshift-bin-\textit{in}dependent color cut used to remove stars from the DESI LRG targets in \citet{Zhou_2023}.}
    \label{fig:stargalcolorcuts}
\end{figure*}

\subsection{Morphology vs. color}
The \extmash{} star-galaxy separator is a morphological classifier, which primarily relies on an object's inferred shape to characterize it as a star vs. the extended morphology of a galaxy. This requires both a small and well-defined PSF, such that extended sources can be differentiated from point sources and the effects of blending are reduced. This imparts a dependence of the classifier on the PSF and its errors, as well as on the density of objects in the field and the ability to deblend them. However, it is quite robust to errors in the photometric colors of objects.

Another common approach for star-galaxy separation is the use of color cuts with NIR data. 
Galaxies emit more infrared photons than typical stars owing to the abundance of low-mass cool stars (in late-type galaxies) and/or the re-processing of starlight through dust (in early-type galaxies).
H$^-$ has a minimum opacity at $\sim$1.6$\mu$m (restframe), resulting in a maximum in the spectral energy distribution (SED) of stars that are $\gtrsim 1$ Myr \citep{Sawicki_2002}. When galaxies populated by such stars are redshifted, the peak shifts further into the infrared. 

Approaches that integrate both color and morphological information (e.g. \cite{Cabayol_2018}) or operate directly on images like \cite{Zhang_2024} show promise, but can also result in more complicated systematic dependencies that are hard to characterize. Machine learning approaches trained to predict star-galaxy separation on a narrow subset of the data may not accurately transfer to the full data set if the training data is not representative in terms of data properties, including the types and relative frequencies of objects. Such \textit{domain shift} is especially likely for generic star-galaxy classifiers because stellar densities vary so strongly across the sky, such that the implicit prior class-probabilities learned during training are unlikely to hold when applied to another part of the footprint.

\citet{Sevilla_Noarbe_2021} perform several tests of the \extmash{} classifier on Y3 Gold objects, comparing against ``truth" samples of stars that are obtained using (1) Near infrared (NIR) data from the Vista Hemisphere Survey (VHS) for objects with matches (limited to $\textit{i} \lesssim 21$) and (2) to HSC-SSP DR2 objects overlapping DES objects in the SN-X3 field, which also uses a morphological classifier but has a smaller PSF \cite{Aihara_2019}. 
From the VHS comparison they find significant contamination at the bright end from binary stars (up to 30\%), which they suggest can be mitigated for cosmology samples by removing the brightest samples or using NIR data for objects that have matches. This motivates the bright limit in Eq.~\ref{eq:maglim_select}, though without additional cuts Fig.~9 suggests that an \extmash{}=3 cut would result in a residual $\sim6\%$ stellar contamination rate for the brightest objects. 

Their comparison to HSC in the SN-X3 field suggests a residual contamination rate of $<2\%$ for the faintest galaxies. We note that since stellar contamination rates are highly dependent on the ratio of galaxies to stars in the field, these values should not be taken as directly generalizable to the entire \maglim{} sample, but they confirm that additional color-based selection can likely mitigate residual stellar contamination in the sample, and give a very rough guide as to what levels we might expect.

\subsection{NIR data from \unWISE{} via \decals{}}
\unWISE{} \citep{2019ApJS..240...30S} is a public data release of data from the Wide-field Infrared Survey Explorer \citep[WISE;][]{2010AJ....140.1868W} that covers the full sky. \WISE{} has four infrared bands, ranging from 2.8 to 28 $\mu m$, but the signal-to-noise is such that we use only the bluest band, W1, centered at $3.4\mu m$. This makes the W1 color increasingly discriminating between stars and galaxies up to a redshift of $z\sim 1.1$, provided there is enough SNR.
We take advantage of the fact that DR9 of \decals{} has already computed infrared photometry for all of its sources for DESI targeting purposes, and covers the entire DES footprint \cite{DESI:2018ymu}. 
They use a ``forced photometry" approach, where the optical source positions and profiles are used to model the infrared flux on the \WISE{} images, respecting the \WISE{} point spread function and noise model. 
This allows measurements of the infrared fluxes even at very low (or zero) signal-to-noise, as described in \citet{Lang_2016}.

We perform object-by-object matching of the \maglim{} catalog with \decals{} to get the associated \unWISE{} fluxes. We only consider the \unWISE{} fluxes usable for star-galaxy separation if we find a match in DR9 and that object would also not be removed when applying the DESI LRG mask, which masks a larger (magnitude-dependent) radius around bright stars than the default optical star mask to account for the large ($\sim6$ arcsec) \wise{} PSF (see \citet{Zhou_2023}). This is necessary to exclude regions where W1 photometry may be unreliable due to stellar wings, diffraction spikes, etc.

We find matches for 99.67\% of the DES \maglim{} objects in \decals{} via simple RA/DEC matching (using 1 arcsec match radius). Visual inspection of a random subset of these shows that most of the objects without matches appear to be object identification artifacts, with no galaxies observed at the given RA/DEC in the DES photometry, though we estimate that it is possible that up to $\sim30\%$ could be actual galaxies where the \decals{} catalog inferred a different deblending solution. Inspecting regions where there are higher than average false detections indicates some by-eye correlation with diffuse galactic cirrus.

Of the matched objects, $4.51\%$ fall within regions that would be removed via the DESI LRG mask, so we do not trust their \unWISE{} fluxes. 
We then define NIR color-based selections for each redshift bin to identify residual stellar contamination in the remaining $~95.2\%$ of \maglim{} objects (97.9\% of these objects have W1 SNR $> 3$).

\begin{figure*}
    \centering
    \includegraphics[width=1\linewidth]{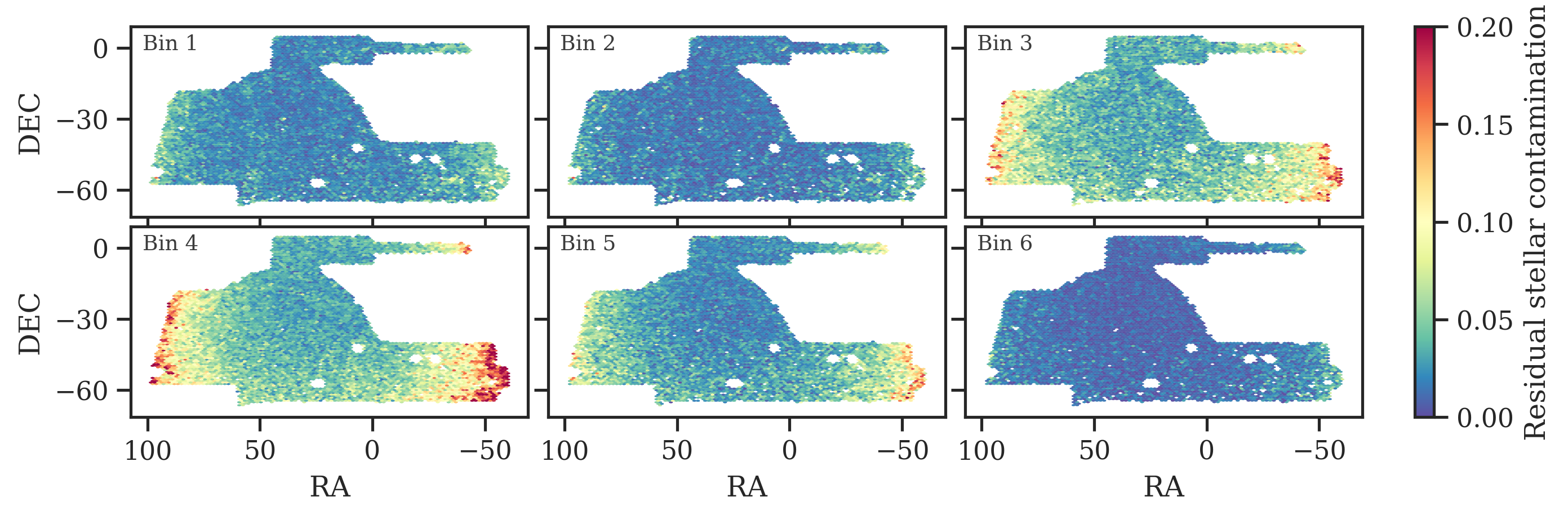}
    \caption{Fraction of \maglim{} galaxies classified as stars via the redshift-bin-optimized NIR classifier. Note the classifier is purely color-based and contains no information about (RA, DEC), but clearly selects objects that trace the density of Milky Way stars. The strong spatial and redshift-bin dependence of contamination shows how characterizations of a single residual stellar contamination rate for galaxy samples should be treated with caution and not assumed to apply directly to subsets of the data. Only pixels with at least 100 objects are shown.}
    \label{fig:starsfootprint}
\end{figure*}

\subsection{Identifying stellar contamination}
In the ($r-z$, $z-W1$) plane, stars occupy a compact locus at small $z-W1$, while galaxies span a broader region at larger $z-W1$. Within the galaxy locus, older quiescent galaxies tend toward redder $r-z$ while younger, star-forming galaxies at the same redshift trend bluer. This accounts for the two main peaks (yellow contours) seen in the higher redshift bins of the \maglim{} sample in Fig.~\ref{fig:stargalcolorcuts}.

To remove stars for the DESI LRG targets, \citet{DESI:2022gle} defined a single linear separation in the ($r - z$, $z - W1$) plane, similar to the approach in SDSS wherein a linear cut in ($r - i$, $r - W1$) was used \citep{Prakash_2016}. While effective, a global cut applied to the full sample is suboptimal ---
galaxies are already grouped into different redshift bins according to their color (via their \dnf{}-inferred \zmean), which means that objects in each bin populate different regions of color space, and thus we can significantly improve upon a redshift-bin-agnostic separation between stars and galaxies by  accounting for how the star-galaxy boundary shifts across redshift bins.

We proceed as follows:
First, we plot the distribution of objects in each redshift bin in the ($r - z$, $z - W1$) plane,\footnote{We use DES photometry for all bands except W1, which directly comes from \decals{}. The per-object agreement between \decals{} DR9 and DES Y3 photometry is quite high, with e.g. $\langle\Delta (r-z)\rangle\lesssim0.03$, but they are not identical. Either could be used for the star classification, provided the same colors are used to both define and apply the classifier.} plotting iso-contours corresponding to the $\{32, 68, 90, 95, 97, 99\}^{\rm th}$ \%iles. Since stellar contamination is expected to be $\mathcal{O}(1\%)$, it is important to have contours out to the tails of the distribution so as to identify the peak in color-space that comes from stars. 

Second, we take as a starting line the relation used for star-galaxy separation of LRGs in \citep{DESI:2022gle} and then adjust it within each redshift bin to lie along the "valley" between the peaks from galaxies (large $z - W1$) and that of stars (small $z - W1$). We allow for a piecewise linear fit with a single kink, which does a good job of isolating the stellar locus without adding many degrees of freedom. This is particularly helpful at higher redshift to avoid removing a tail of galaxies that extends down to small $z - W1$, and would be removed with the single linear cut such as applied in \citet{DESI:2022gle}.

The resultant densities are shown in Fig.~\ref{fig:stargalcolorcuts}. We classify objects below the $W1 - z$ threshold as stars and remove them from the sample, finding residual stellar contamination fractions of \{2.4, 1.8, 4.9, 5.5, 3.4, 1.3\}\%  across the Y3 \maglim{} bins, and 3.3\% across the full sample. 

The density of these objects across the footprint is plotted in Fig.~\ref{fig:starsfootprint}, and shows a very similar dependence on galactic latitude as expected for true stars. We combine all stars into pixels of \nside{}=64 in order to get good enough SNR and compute a (pixel-area-weighted) Pearson correlation coefficient with the DES Y3 stellar density map of 0.928. 

To illustrate the impact on the density field, we plot the observed overdensity with (blue) and without (orange) our bin-optimized star-galaxy cuts in Fig.~\ref{fig:dens_vs_stars}. Removing the residual stellar contamination in this way all but removes the otherwise significant trends of the observed density field with stellar density, resulting in a considerably cleaner sample.

\subsection{Comparison of NIR and HSC classification}
We further match our $\maglim{}\times\decals{}$ catalog with the HSC PDR3 wide field catalog \cite{Aihara_2022},\footnote{We remove objects where \texttt{\{band\}\_cmodel\_flux}=0 or any of the following flags are \texttt{TRUE}: \texttt{\{band\}\_pixelflags\_interpolatedcenter}, \texttt{\{band\}\_pixelflags\_saturatedcenter}, \texttt{\{band\}\_extendedness\_flag}, for $\texttt{band} \in \{g,r,i,z\}$.} finding 648615 objects over $\sim 300$ deg$^2$ in the region roughly spanned by RA $\in [-30, 40]$, DEC $\in [-7, 3]$ with good photometry across all three surveys. Using the morphological star-galaxy classifier (${i_{\rm extendedness}}=1$ \cite{Li_2022}), we compare how \maglim{} galaxies are classified using both the NIR and HSC classifiers in Table~\ref{tab:triplematch_classify}.

\begin{figure*}
    \centering
    \includegraphics[width=.8\linewidth]{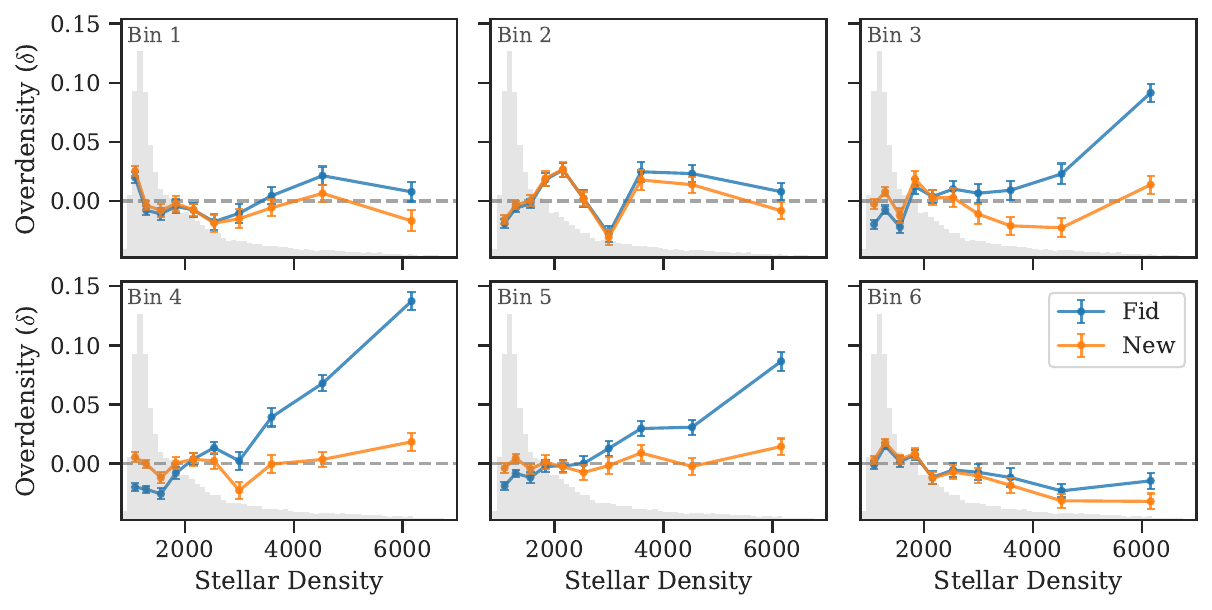}
    \caption{Observed density contrast vs. Gaia stellar density for fiducial \maglim{} galaxies (blue) and after cutting objects defined as stars with the bin-optimized NIR star-galaxy classifier described in this work. We show the  mean overdensity in bins of stellar density, with bootstrapped error bars. Bin widths are defined as a weighted average of equal-width and equal-count bin widths, so as to have resolution in both the bulk and the tail of the distribution of stellar densities (gray histogram). The star-galaxy cuts described here all but eliminate the dependence of the overdensity on stellar density.}
    \label{fig:dens_vs_stars}
\end{figure*}

\begin{table}
    \centering
    \begin{tabular}{c|cc|c}
         &  HSC Star&  HSC Galaxy& Total\\\hline
         NIR Star&  0.9\%&  1.8\%& 2.7\%\\
         NIR Galaxy&  0.2\%&  97.1\%& 97.3\%\\\hline
         Total&  1.1\%&  98.9\%& 100\%\\
    \end{tabular}
    \caption{Classification rates for  \maglim{} galaxies that are cross-matched with objects in \decals{} and HSC.  Note that both HSC and DES classifiers are based on morphology and so will have more correlated errors (e.g. due to deblending errors). }
    \label{tab:triplematch_classify}
\end{table}

\begin{figure*}
    \centering
    \includegraphics[width=1\linewidth]{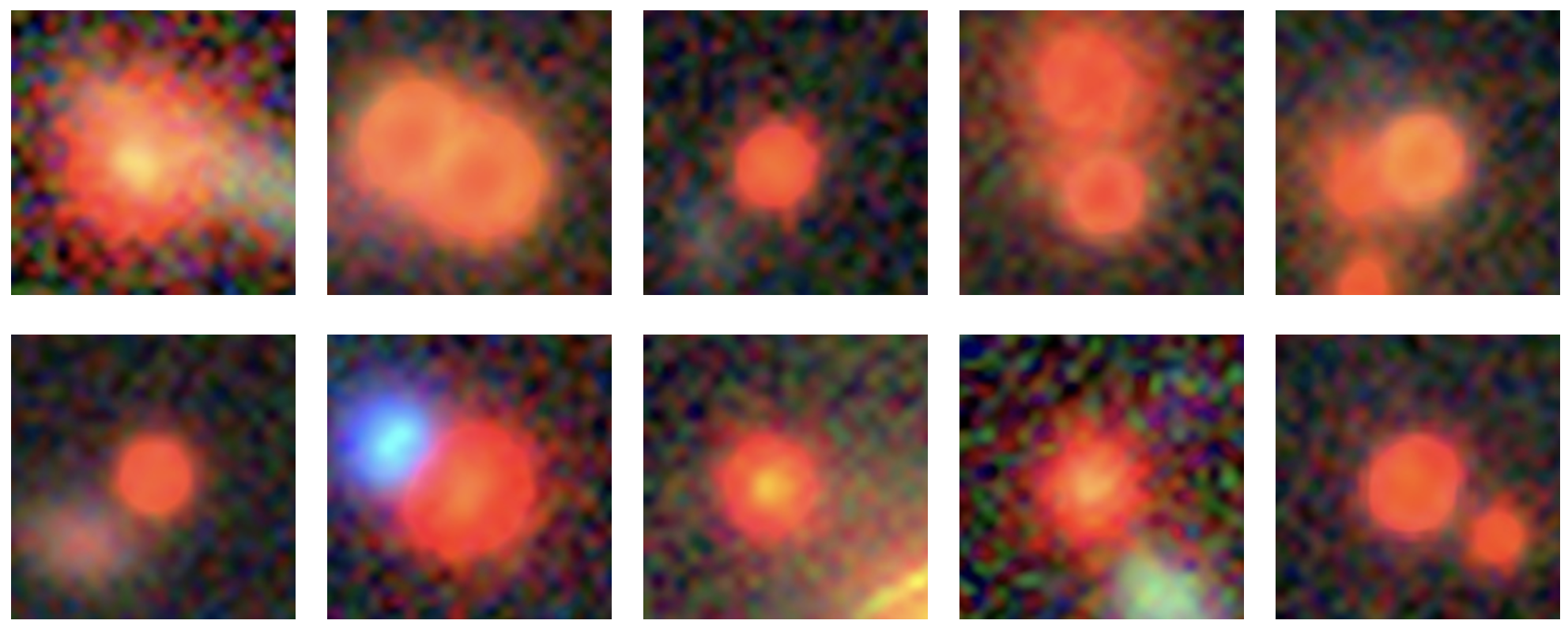}
    \caption{HSC images for a random subset of 10 objects that are classified as galaxies by both HSC and DES morphological classifiers but stars by our NIR cuts in bin 4, demonstrating the impact of blends on the morphological classifier. Each cutout has a side length of 4".}
    \label{fig:cutouts}
\end{figure*}

There is agreement for 97.1\% of the sample, with both NIR and HSC classifiers confirming the DES \texttt{EXTMASH} galaxy classifications. However, the remaining 2.9\% are classified as stars by at least one of these; with NIR estimating 2.7\% stellar contamination and HSC estimating 1.1\%. Almost all ($\sim80\%$) of the HSC-classified stars are also classified as such via the redshift-optimized NIR cuts. 
Note that the HSC and DES classifiers are both morphological and so we expect their (mis)classification rates to be correlated. E.g.  any residual stellar contamination in \maglim{} that is due to deblending issues, where multiple close objects are instead classified as a single extended object, may pose challenges to both DES and HSC classifiers, though the smaller PSF of HSC should resolve some of these misclassifications. We therefore expect the HSC classifier to be an underestimate of the true residual stellar contamination. 
This is supported by visual inspection of a random subset of objects classified as stars via NIR but galaxies by HSC, which  reveals a large fraction of blends. A random set of ten of these are shown in Fig.~\ref{fig:cutouts}.

An alternative explanation is that in crowded regions, the model fit from \texttt{Tractor}\footnote{\url{https://thetractor.readthedocs.io/}} \cite{2016ascl.soft04008L} mis-distributes the \wise{} flux across objects, resulting in an over- or (more relevant here) under-estimate of the true \wise{} flux for the object in question. If for some unknown reason the \wise{} flux were to be systematically under-allocated to DES \maglim{}-selected galaxies but over-allocated to their close neighbors, then this \textit{could} result in false star detections by the NIR classifier.

We therefore estimate a very conservative upper bound of the false NIR-stars by computing an upper limit of the \wise{} flux for each NIR star in the DES$\times \decals{}\times$HSC sample by assigning it the sum of the \wise{} flux for \textit{all} its neighbors identified in the \decals{} data (not just neighbors in \maglim{}), defined as objects within a 6 arcsec radius (i.e. 7$\times$ the area of each cutout in Fig.~\ref{fig:cutouts}). These NIR stars have an average (std) of 2.2 (1.3) neighbors contributing their fluxes, with the mean new \wise{} $\sim2.8\times$ higher when neighbors' fluxes are added. 
This reduces the estimated NIR stellar fraction from 2.7\% to $1.4\%$. This represents a rather  conservative floor for stellar contamination rate of this matched sample, as we've roughly tripled the assigned \wise{} flux of the average object. That the true rate is closer to our initial estimate of 2.7\% is further supported by the absence of a strong negative correlation between galaxy density and stellar density after removing the NIR stars (orange curve in Fig.~\ref{fig:dens_vs_stars}).

\begin{figure*}
    \centering
    \includegraphics[width=1\linewidth]{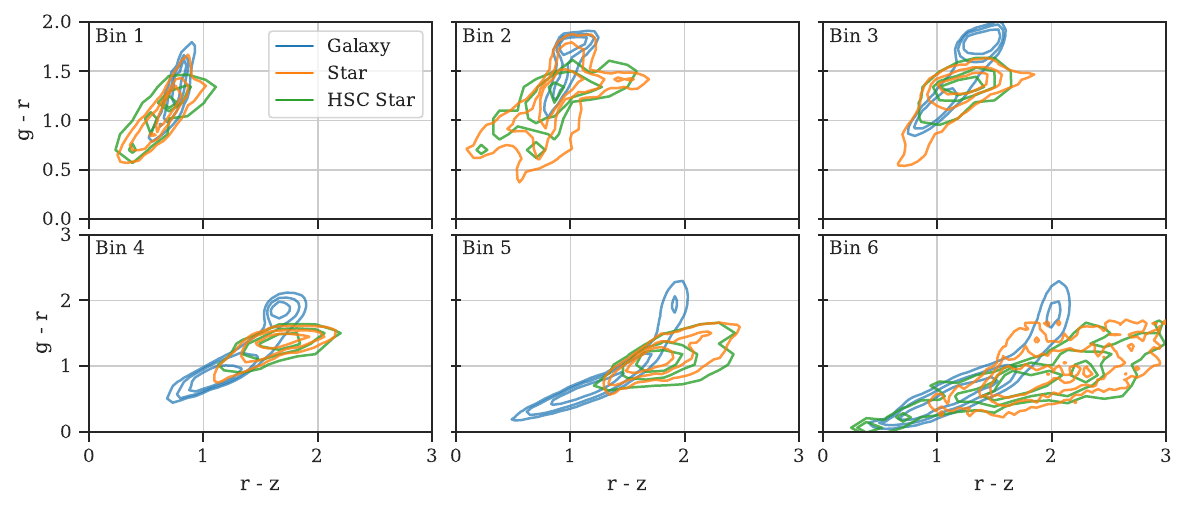}
    \caption{DES color distribution of objects classified as galaxies (blue) vs. stars (orange) using the redshift-bin-optimized NIR star-galaxy classifier described in this work, as well as objects classified confidently as stars by the HSC classifier where DES and HSC overlap. Curves indicate the \{25, 50, 75\}$^{\rm th}$ percentile contour lines. The color distribution of the NIR stars aligns much more closely with that of the HSC stars than with the galaxies, suggesting that these are indeed correctly classified as such and not due to W1 flux mismodeling.}
    \label{fig:hscstar_colors}
\end{figure*}

As a final check, we also inspect the DES color distribution of NIR-galaxies (blue) vs. NIR-stars (orange) and HSC-stars (green) in  Fig.~\ref{fig:hscstar_colors}. We show the \{25, 50, 75\}$^{\rm th}$ percentiles of each distribution. While the contours have some minor discreteness effects because of the limited number of objects (particularly for the HSC stars, which only use the triple-matched sample), it is apparent that the color distribution of the NIR-stars follows much more closely that of the HSC stars, despite the latter being a purely morphological classifier. This supports the assessment that the NIR-stars are likely to be correctly classified and it is the morphological classifiers that have failed for these objects.

\subsection{Comparison with no redshift bin optimization}
Finally, we compare our results to the case with the generic, non-redshift-bin-optimized star-galaxy cut used for the DESI LRG targets in \citet{Zhou_2023}.\footnote{Note that the DESI LRG targets populate a slightly different region in color-space, because of their luminosity cut (Eq.~2d in \cite{Zhou_2023}) which keeps only objects that are bright in W1.} This corresponds to the dashed gray line in Fig.~\ref{fig:stargalcolorcuts}. We again use the triple-matched sample to compare against HSC-classified stars. While 2.6\% of the triple-matched sample are classified as NIR stars in the bin-optimized approach, this jumps to 8.6\% when applying the non-optimized color cut, representing a far more aggressive cut. Only 1.0\% of these additional objects are classified as stars with the HSC criterion, as compared to 32\% of the bin-optimal NIR-classified stars (Table~\ref{tab:triplematch_classify}), indicating that the vast majority of the objects removed are galaxies. The bin-optimized approach enables a more precise sculpting of the galaxy sample that preserves number density without introducing additional stellar contamination. 

An added benefit of the bin-optimized cuts shown in Fig.~\ref{fig:stargalcolorcuts} as opposed to using the same cut as \citet{Zhou_2023} is that the bin-optimized cuts have reduced potential for introducing angular systematics. Each cut applied to the sample creates a sharp boundary in color space that presents an additional way for photometric errors to translate into variations in observed galaxy density as objects move in and out of the selection.
Locally, the response of the galaxy counts to photometric calibration errors is given by the gradient of $N_g$ across the selection boundary:\footnote{Note that this is the gradient of the number of galaxies \textit{in the sample}, i.e. it's large in regions of color-space with a selection that cuts through a high density of galaxies. This is not to be confused with the gradient of the density of all sources.}
\be\label{eq:deltaNboundary}
\delta N \approx \nabla_\mathbf{m} N_g \cdot \delta \mathbf{m},
\ee
evaluated pointwise along the boundary, where $\delta \mathbf{m}$ is the photometric calibration error in each band. The total response is obtained by integrating this contribution over the full boundary of the sample selection.

Because the redshift-optimized cuts in this work are applied in the tails of the distribution where $\nabla_\mathbf{m} N_g$ across the selection boundary is small, the sample is less susceptible to systematic errors in the measured W1 flux than the DESI-style cut, which slices through higher-density regions of color space for this sample (Fig.~\ref{fig:stargalcolorcuts}).
Using finite differences, we measure the susceptibility of the full \maglim{} sample to a spatial calibration error $\delta W1 = 0.01$ for each of the two different star cuts, finding
\begin{equation*}
    \left(\frac{\delta N_g}{N_g}\right)_{\rm DESI\ cut} = 2.6\times10^{-3}, \quad \left(\frac{\delta N_g}{N_g}\right)_{\rm zopt} = 2.0\times10^{-4},
\end{equation*}
an order of magnitude smaller sensitivity for the redshift-bin-optimized cut.
The difference is most stark for the first redshift bin, where we find sensitivities of $9.1\times10^{-3}$ vs. $3.0\times10^{-4}$, respectively. As seen in Fig.~\ref{fig:stargalcolorcuts}, the dashed gray curve cuts through a much higher density region for Bin 1 than does the bin-optimized red curve, such that the same error in W1 would result in a much larger flux of galaxies in or out of the sample.  

We note that defining selections so as to minimize Eq.~\ref{eq:deltaNboundary}, i.e. minimizing the number of cuts with a large gradient across them, may be a promising approach to make galaxy samples generically more robust to angular systematics. At a minimum, using knowledge of the sample cuts to \textit{a priori} identify the directions in color space to which the galaxy density is most sensitive to photometric errors 
(i.e. the cuts with the largest gradient $\nabla_\mathbf{m} N_g$ across their boundaries)
would provide well-motivated linear combinations for band-dependent survey property maps as a more principled basis of systematic templates to feed to regression algorithms for estimating contamination. 

\section{Conclusion}

In this work, we address the problem of additive systematics arising from stellar contamination in photometric galaxy samples used for cosmological analyses. We first derived optimal galaxy weights for mitigating both additive and multiplicative contamination to the density field at the map level, extending the standard weighting framework commonly employed in large-scale structure analyses. 
We demonstrated how these optimal weights improve the fidelity of the overdensity field and the two-point angular correlation function as compared to current weighting schemes, in particular pointing out how other weighting schemes will unintuitively result in inconsistent linear bias estimates when subsets of the footprint are used, such as for cross-correlation studies.

While these weights are formally optimal given known spatial templates and contamination fraction, we argue that direct removal of stellar contaminants from the sample is preferable in practice, as it mitigates suppression from the wrong integral constraint and avoids the complex procedures (and additional mode nulling) that comes from fitting for both additive and multiplicative systematics that share the same (and approximate) spatial templates.

We proposed an approach to star-galaxy separation for galaxy samples that are binned by photometric redshift. Because the color distribution of objects changes strongly between redshift bins, so does the optimal boundary separating stars and galaxies. This can be viewed as reducing the domain shift between the training set on which the star-galaxy separation is defined and the test set on which it's applied. 
This bin-optimized color-based star-galaxy classifier is complementary to morphological classifiers, whose performance can be sensitive to PSF modeling and source blending.

We demonstrated this approach on the DES Y3 \maglim{} lens sample, leveraging forced NIR photometry from unWISE obtained through cross-matching with \decals{} DR9 to construct redshift-bin-optimized color cuts. We estimate residual stellar contamination rates of \{2.4, 1.8, 4.9, 5.5, 3.4, 1.3\}\% across the Y3 \maglim{} photometric redshift bins (3.3\% in aggregate). We compared overlapping objects to their morphological star-galaxy classification from HSC PDR3, finding that a third of the residual stars are classified as point-like with HSC's smaller PSF, and the others have a large fraction of visible blends. We tested the robustness of the results to mismodeling of source WISE fluxes, placing a conservative lower limit of at least half of the NIR-stars being true stars. Finally, we compared the redshift-optimized cuts proposed here to the global star-galaxy cut used for DESI LRGs in \cite{Zhou_2023} from which it was adapted, finding that the additional objects removed by the global cut are dominated by true galaxies, and such a cut increases the sensitivity of the sample to systematics from the W1 measurements.

This bin-optimized star-galaxy separation approach has already been used to estimate and correct for residual stellar contamination in the DES Y6 BAO sample \cite{Mena_Fern_ndez_2024}, and incorporated into the fiducial DES Y6 \maglimpp{} lens sample \cite{weaverdyck2026darkenergysurveyyear} for \threetimestwo{} pt analyses. 
Looking ahead, this approach is applicable to lens samples from upcoming photometric surveys such as the LSST, Euclid, and the \textit{Roman Space Telescope}. Effective application ideally requires the combination of optical and near-to-mid infrared photometry at comparable depths -- e.g. to achieve color separation as precise as that achieved here for the $i\lesssim22.2$ \maglim{} sample, the prototypical LSST Gold sample ($i\lesssim25.1$) would require $W1$ depths roughly three magnitudes deeper.

More research is needed to determine the performance of NIR color-based star-galaxy separation when using deep observations with $\lambda \lesssim 2 \mu$m, such as from a full-sky NIR survey from Roman as described in \cite{han2026pathallskysurveyroman}. While bluer than $W1$, the exquisite 0.1 arcsec resolution means that such a survey would significantly augment both color-based and morphological star-galaxy classification for LSST. 
Without sufficiently deep near-to-mid-infrared counterparts across the full footprint, careful application of the weighting scheme of equation~(\ref{eq:optweight}) along with well-characterized models of the additive and multiplicative contamination $\eta_{\rm add}$, $f_\mathrm{add}$, and $f_\mathrm{mult}$ provides a robust way to correct for residual contamination. 

More broadly, we advocate for aggressive removal of potential sources of additive systematics at the catalog level, while doing so judiciously to minimize the risk of such cuts imparting systematic errors from greater photometric sensitivity. The approach detailed here offers a path toward taming the dominant source of additive systematics in photometric galaxy surveys, and a primary systematic in galaxy clustering measurements.

\section*{Data availability Statement}
The original DES Y3 \maglim{} data underlying this article are available in the Dark Energy Survey Data Management platform at \url{https://des.ncsa.illinois.edu/releases}. A matched catalog including W1 fluxes and stellar classification will be made available upon publication.

\begin{acknowledgments}

We thank Rongpu Zhou, Alex Drlica-Wagner, Jack Elvin-Poole and Mart\'in Rodr\'iguez-Monroy for helpful conversations.
NJW is supported by the Chamberlain Fellowship at Lawrence Berkeley National Laboratory.

The analysis made use of the software tools {\sc SciPy}~\citep{Jones:2001}, {\sc NumPy}~\citep{Oliphant:2006}, {\sc Astropy} ~\citep{astropy:2013, astropy:2018}, \healpix{} \cite{Gorski_2005}, and {\sc Matplotlib}~\citep{Hunter:2007}.

This research used resources of the National Energy Research Scientific Computing Center (NERSC), a Department of Energy User Facility.

\end{acknowledgments}



\appendix
\section{Additive and Multiplicative Errors}
\label{sec:app1}

One can generically model how such systematics affect the density maps, where the observed density of objects in a given volume element is \cite{Weaverdyck:2020mff}:

\begin{align}
N_{\rm obs}(\xvec) &= N_{\rm true}(\xvec) \left(1 + \fmultp(\xvec)\right) + N_{\rm add}(\xvec)\\
    &= N_{\rm true}(\xvec) +  N_{\rm mult}(\xvec) + N_{\rm add}(\xvec)
\end{align}
with contributions from both the true galaxy density modulated by a generic, position-dependent function $1+\fmultp(\xvec)$ (characterizing multiplicative systematics), and unwanted objects due to additive contaminants $N_{\rm add}(\xvec)$, which are independent of the true galaxy field.\footnote{Note we can write the number of additional objects due to multiplicative systematics as $N_{\rm mult}(\xvec) = \fmultp (\xvec) N_{\rm true}(\xvec)$.
This assumes that additive artifacts aren't modulated by multiplicative systematics. This may or may not be true, but in any case as interlopers their modulation is unlikely to be described by the same selection function as the true galaxies, and the effect will be of order $\fmult \fadd \lesssim \mathcal{O}(10^{-3})$ at most for typical contamination rates in modern cosmological surveys.}

Hereafter we drop $(\xvec)$ for clarity, but note that all quantities are position-dependent (typically corresponding to a given pixel and redshift bin) unless noted otherwise, or if they contain an overhead bar such as $\overline{f}$, indicating the spatial average for the given sample.
Defining the density contrast as $\delta = N/\overline{N} - 1$, we have

\begin{align}
\deltasub{obs} &= \frac{\Nsub{true} + \Nsub{add} + \Nsub{mult}}{\overline{N}_{\rm obs}} - 1 \\
                &= \left(1 - \frac{\overline{N}_{\rm add}}{\overline{N}_{\rm obs}}\right) \deltasub{true} + \fmult + \fadd + \fmult \deltasub{true}, 
\end{align}
where we have defined the systematic functions:
\begin{align}
\fmult &\equiv \frac{\fmultp\overline{N}_{\rm true} - \overline{\fmultp N_{\rm true}}}{\overline{N}_{\rm obs}}\\
    &= \frac{ \overline{N}_{\rm true} \left(N_{\rm mult}/N_{\rm true}\right)  - \overline{N}_{\rm mult}}{\overline{N}_{\rm obs}}
\end{align}
and
\be
\fadd \equiv \frac{N_{\rm add} - \overline{N}_{\rm add}}{\overline{N}_{\rm obs}},
\ee
which are analogous to the zero-centered systematics in Sec.~4.2 of \cite{Weaverdyck:2020mff}, but extended to explicitly account for additive systematics. 

These are mean-zero systematic fields, which are typically estimated by assuming their spatial dependence is traced by templates ($f_X\sim \mathcal{F}(t_0, t_1,...t_{\rm \Ntpl})$) given by maps of survey observing properties, stellar density, dust, etc., and fitting the observed overdensity to these templates.

Any observed density fluctuations that are well modeled by these templates are assumed to be spurious:

\be
\deltasub{obs} \sim \mathcal{F}(t_0, t_1, ... t_{N_{\rm tpl}}, \deltasub{true}) + \deltasub{true} + \epsilon
\ee

This general dependence of $\mathcal{F}$ on $\deltasub{true}$ is typically further specified by whether the systematics treatment method assumes the contamination is additive or multiplicative. Regression or some other form of supervised learning is used to estimate the best-fit parameters of $\mathcal{F}$ given $\deltasub{obs}$ and the templates (with tradeoffs between the freedom given to the model $\mathcal{F}$ and the unintentional removal of true LSS fluctuations), and this used to generate galaxy weights or otherwise correct the data used for cosmological inference.\footnote{Refs. ~\citep{Johnston_2021, Yan_2025} propose an approach that is somewhat different from most of the supervised learning approaches using templates. In the ``organized randoms" approach, they construct a self-organizing map (SOM) to essentially group pixels on the sky into $N_{\rm grp}$ groups according to their proximity in the $N_{\rm tpl}$- dimensional template space. Pixels in each group are assumed to have the same multiplicative contamination, and deviations of each group from the global mean removed by modulating the randoms. This amounts to effectively fitting $N_{\rm grp}$ parameters using the observed overdensity, but is less parametric in the prescription of how $\mathcal{F}$ depends on $t_i$. However it comes at the cost of being more discrete in the corrections and some non-obvious dependence on the choice of distance metric adopted for the SOM (e.g. it is not obvious how to relate the ``distance" between two seeing values vs. that between two stellar density values when creating the SOM, except for their relative impact on the observed density which isn't included in training the SOM).}

\section{Comparison to \citet{Crocce_2015}}
Appendix D of \citet{Crocce_2015} derives the impact of stellar contamination on the observed galaxy contrast and two-point correlation function, but get slightly different results. For example, when setting $\fmult=0$ and noting that $\fadd = \eta_{\rm add} \left(\frac{N_{\rm add} - \overline{N}_{\rm add}}{\overline{N}_{\rm add}}\right) =\eta_{\rm add}\deltasub{add}$, our Eq.~\ref{eq:twopt_obs} reduces to

\begin{align}
\langle\deltasub{obs}\deltasub{obs}\rangle &= \left(1 - \eta_{\rm add}\right)^2 \langle \deltasub{true}\deltasub{true}\rangle + 
\langle\fadd\fadd\rangle \\
&= \left(1 - \eta_{\rm add}\right)^2 \langle \deltasub{true}\deltasub{true}\rangle + \eta_{\rm add}^2\langle \deltasub{add}\deltasub{add}\rangle \\
&= \left(1 - \eta_{\rm add}\right)^2 w_{\rm true} + \eta_{\rm add}^2 w_{\rm add}, \label{eq:twopt_addonly}
\end{align}
where $w_X$ is the two-point correlation function of field $X$.
This differs from Eq.~41 in \cite{Crocce_2015} because of a subtle error in their derivation, wherein they define $f_{st}$ as the mean stellar contamination, normalized to the \textit{observed} density (i.e. equivalent to our\footnote{Note also their $\langle N_{st,S}F(S)\rangle$ is equal to our $\bar{N}_{\rm add}$.} $\eta_{\rm add} = \bar{N}_{\rm add}/\bar{N}_{\rm obs}$), but it is at times treated as also equivalent to $\bar{N}_{\rm add}/\bar{N}_{\rm true}$ (for instance in their Eqs.~(35) and (37)). These definitions collide and a spurious term is picked up such that Eq.~(39) and onward are in error.


\bibliography{refs_truncated} 



\label{lastpage}
\end{document}